\documentstyle[a4wide,11pt,epsf]{article}

\begin{document}
\title{Cosmic string network evolution in arbitrary Friedmann--Lema\^{\i}tre 
models}
\author{Carsten van de Bruck\\
        Institut f\"ur Astrophysik und Extraterrestrische Forschung\\
        Auf dem H\"ugel 71\\
        53121 Bonn, Germany}
\maketitle

\begin{abstract}
We use the modified ``one--scale'' model by Martins \& Shellard to investigate 
the evolution of a GUT long cosmic string network in general 
Friedmann--Lema\^{\i}tre models. Four representative cosmological models are 
used to show that in general there is no scaling solution. The implications 
for structure formation in these models are briefly discussed. 
\end{abstract}

\section{Introduction}
Cosmic strings might be responsible for structure formation in the universe. 
The theory predicts three mechanism for structure formation: wake 
formation by fast moving long strings, accretion of matter by cosmic 
string loops and filamentary accretion by slow moving long strings. 
Which of these mechanism is important depends on the evolution of the cosmic 
string network (for a review see \cite{VileShell},\cite{HindmarshKibble}).

Up to now, the cosmic string scenario of structure formation 
has been investigated only in the Einstein--de Sitter model, in which the 
cosmic string network reaches a scaling solution, i.e. the typical 
length scale of the network scales with the Hubble radius. Investigations 
in open universes assumed such a scaling solution a priori \cite{Ferreira}.
Recent work indicates that, if the network reaches scaling, the angular 
power spectrum $C_l$ and the COBE normalised matter power spectrum doesn't 
fit the observation \cite{Pen}. 

However, in recent years it became more and more obvious, that the 
Einstein--de Sitter model is in conflict with some astronomical observations, 
namely the age of the oldest stars \cite{Hogan}, the baryonic content 
of X--Ray clusters \cite{White} and the line distribution of hydrogen 
absorbers in the Lyman--$\alpha$--forest \cite{Priester}. 
Therefore, it is necessary to investigate the 
evolution of a cosmic string network in more general Friedmann--Lema\^{\i}tre 
models. The first quantitatve discussion of the evolution of a cosmic 
string network in open models was given by Martins \cite{Martins}. In this 
paper we extend his analysis and discuss 
the evolution of a GUT cosmic string network in more general cosmological 
Friedmann--Lema\^{\i}tre models. 
There are several types which are interesting in modern cosmology. 
The flat models (some of them could be produced in an inflationary epoch, 
but not in general) have $\Omega + \lambda = 1$, where 
$\Omega(t)=8 \pi G \rho(t)/(3 H^{2}(t))$ is the density parameter 
and $\lambda(t) = \Lambda / (3 H^{2}(t))$ is the normalized 
cosmological term ($\Lambda$ is the cosmological constant). The open models 
with $\Omega<1$ are favoured by the measurements of cluster masses 
\cite{cluster}, but $\Lambda=0$ was assumed for the resulting 
cosmological model. The third class of interesting cosmological models 
are closed with a loitering phase of slow expansion. The time and the 
duration of the phase of slow expanison depends on the present values 
of $\Omega$ and $\lambda$. Such a model is suggested by the 
Ly$\alpha$--absorption lines distribution in quasar spectra \cite {Priester}. 
They are considered by several authors in the context of structure formation 
\cite{Sahni}.

Numerical simulations are the simplest (and most expensive) ways to study the 
cosmic string network evolution. Other possibilities are analytical models. 
The first model was introduced by Kibble, the ``one--scale'' model 
\cite{Kibble}. In this model, 
the fundamental quantity is a typical length scale $L(t)$, defined by 
\begin{equation}
\rho_{\infty}(t) \equiv \mu / L^{2}(t),
\end{equation}
where $\rho_{\infty}$ is the energy density in long strings and $\mu$ is 
the mass per unit length on the string (we set $c=1$). With more 
detailed numerical studies it became obvious, that long strings are 
not straight but contain wiggles \cite{Allen}\cite{Bennett}. 
Therefore, Austin, Copeland and Kibble modified the 
``one--scale'' model and introduced two new length scales in order to 
describe these wiggles \cite{Austin}. Another model was introduced by 
Martins \& Shellard \cite{MartinsShellard}, 
in which the RMS velocity of the strings are treated as a fundamental, 
independent quantity. We use this model, to study the evolution of a 
cosmic string network in four representative FL models. This 
velocity--dependent ``one--scale'' model (VDOSM) is briefly described in 
section 2. In section 3 we present our calculations. We discuss the 
results and implications on structure formation and anisotropies of 
the CMBR in section 4. 

\section{The velocity--dependent ``one--scale'' model}
There are only two macroscopic quantities in the VDOSM. The first is the 
energy of a piece of string:
\begin{equation}
E = \mu a(\tau) \int \epsilon d\sigma,
\end{equation}
where $\epsilon$ is the energy per length $\sigma$ on the string. The other 
quantity is the RMS velocity of the (long) string 
\begin{equation} 
v_{\infty}^2 = \frac{\int \dot{{\bf x}}^2 \epsilon d\sigma}
{\int\epsilon d\sigma}.
\end{equation}
The typical length scale $L$ in the network is defined in eq. (1). One 
has to include several phenomenological parameter, the first one is the 
``loop chooping efficiency'' $\tilde{c}$, defined by
\begin{equation}
\left( \frac{d \rho_{\infty}}{dt}\right)_{\rm to \mbox{ } loops} 
= \tilde{c} v_{\infty} \frac{\rho_{\infty}}{L}. 
\end{equation}
Note that in the original ``one--scale'' model the velocity was absorbed in 
the definition of $\tilde{c}$. For our purposes the loop reconnection onto 
long strings is neglegible, as indicated by numerical 
simulations\cite{Allen}\cite{Bennett}. The 
scaling properties depend {\bf not} crucial on the parameter $\tilde{c}$. 

\begin{figure}[htb] 
  \begin{center}
    \leavevmode
    \epsfxsize=15cm
    \epsffile{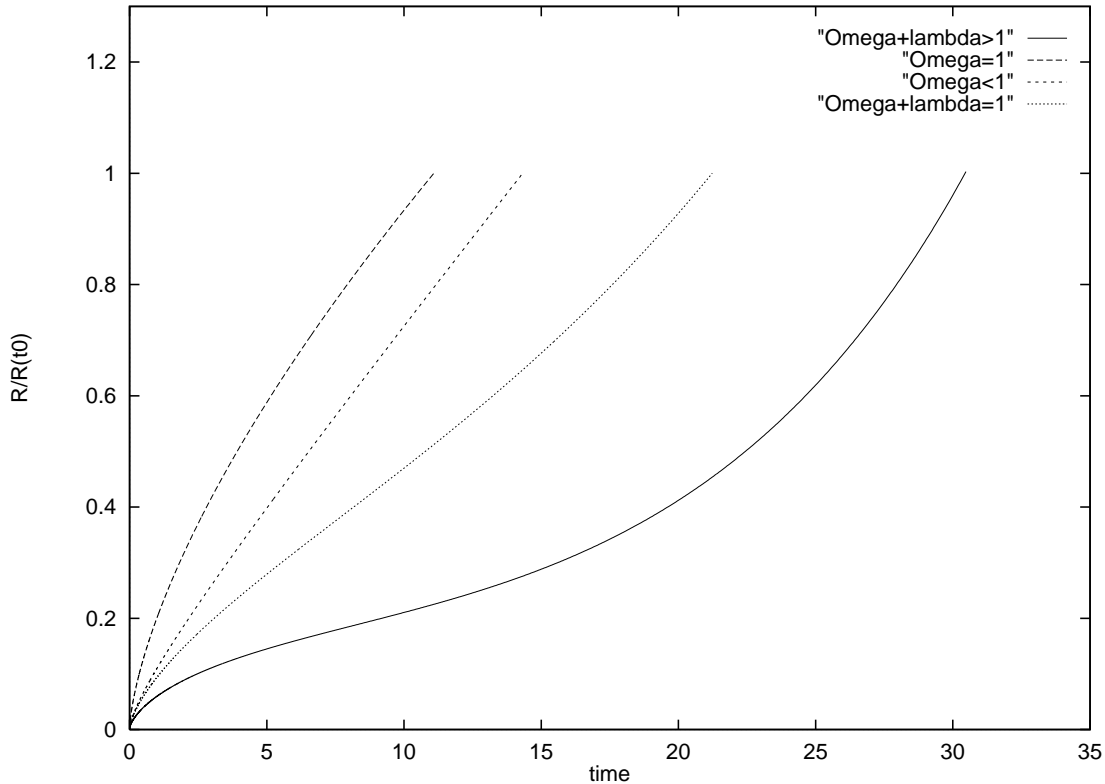}\vspace{0.5cm}
    \caption
    {The scale--factor ($R(t)/R_{0}$) as a function of time (in units of 
$10^{9}$ years) for the four representative 
      models.}
    \label{fi1}
  \end{center}
\end{figure}

Neglecting the effects of frictional forces, the equation for the 
evolution of the length scale $L$ can be obtained from eqs. (2)--(4) and 
is given by: 
\begin{equation}
\frac{dL}{dt} = H L(1+v_{\infty}^2) + \frac{\tilde{c} v_{\infty}}{2}.
\end{equation}
The evolution of the relevant length scale of string loops is described 
by 
\begin{equation}
\frac{dl}{dt} = (1-2v_{l}^2)Hl - \Gamma{'}G \mu v_{l}^6
\end{equation}
where $\Gamma^{'}=8\times 65$. The second term describes the decay of the 
loops due to gravitational radiation. 
Finally, the evolution of the RMS velocity is given by 
\begin{equation}
\frac{dv}{dt} = (1 - v^2)\left(\frac{k}{r} - 2Hv\right),
\end{equation}
where $k$ is another phenomenological parameter that is related to the 
small scale structure on the strings. An appropriate ansatz for it is
\begin{displaymath}
k=\qquad
\left\{
\begin{array}{ll}
1, & 2Hr>\chi \\
\sqrt{2}Hr, & 2Hr < \chi
\end{array}
\right.
\end{displaymath}
Here $r$ is the curvature radius of the string, i.e. $r=L$ for long strings 
and $l=2\pi r$ for loops. $\chi$ is a numerically determined coefficient of 
order unity, see the paper by Martins \& Shellard \cite{MartinsShellard} 
for a complete discussion of this point. 

\section{Computations}
The evolution of the scale factor $R$ is described by the Friedmann 
equation
\begin{equation}
H^2 = \frac{8 \pi G}{3} (\rho_{matter} + \rho_{radiation} + \rho_{\infty}) 
+ \frac{\Lambda}{3} - \frac{K}{R^2}.
\end{equation}
Here $H=\dot{R}/R$ is the Hubble parameter and $R$ is the scale factor. 
$K$ represents the topology of the space and is zero for a flat universe, 
$-1$ for an open universe and $+1$ for a closed universe. $\Lambda$ is the 
cosmological constant. We analyse four representative models, a open 
(hyperbolic) model, a flat model with a cosmological constant, a closed 
universe with a cosmological constant and the (flat) Einstein--de Sitter 
model. The behaviour of the scale factor for these models is plotted in 
Figure 1. These four models represent the interesting class of models in 
modern 
cosmology. The flat model with a cosmological constant could be introduced to 
retain the flatness while lowering $\Omega_{0}$. The open model was 
introduced in favour for a low $\Omega_{0}$ with $\Lambda = 0$. The 
closed model was obtained from the Ly--$\alpha$--forest \cite{Priester}, 
by assuming a constant comoving absorber density and represents the class 
of loitering models, in which the universe undergoes a epoch of slow 
expansion at a redshift about 5.

We solve the equations numerically with the standard Runge--Kutta method. 
Our results are presented in Figures 2 to 6. Our results for the 
Einstein--de Sitter model and for the open model are in agreement with the 
calculations by Martins \cite{Martins}. 

\begin{table}
\begin{center}
\begin{tabular}{|l|l|l|l|l|}
\hline 
Model & K & $\Omega_{0}$ & $\lambda_{0}$ & H$_{0}$/(km/(s$\cdot$ Mpc)) \\
\hline
1     & +1& 0.014 & 1.08 & 90 \\
\hline
2     & 0 & 1.0   & 0.0  & 60 \\
\hline
3     & -1& 0.1   & 0.0  & 60 \\
\hline
4     & 0 & 0.1   & 0.9  & 60 \\
\hline
\end{tabular}
\caption{The four representative cosmological models.}
\end{center}
\end{table}

One can see, that only in the Einstein--de Sitter model the network approaches a scaling 
regime. As pointed out by Martins, this is easy to understand:
In a universe where the scale factor grows as $R\propto t^{s}$ ($s < 1$), 
one finds for in the linear regime 
\begin{equation}
\left(\frac{L}{t}\right )^{2} = \frac{k(k+\tilde{c})}{4s(1-s)}
\end{equation}
and
\begin{equation}
v^2 = \frac{k(1-s)}{s(k+\tilde{c})}.
\end{equation}
In the Einstein--de Sitter model $s$ varies only from $s=1/2$ to $s=2/3$ 
(radiation to matter dominated). In the other models, however, there 
are several other epochs, namely curvature dominated epochs and vacuum 
dominated epochs. In the vacuum dominated epochs, the scale factor grows 
as $R \propto \exp(t)$, therefore $s$ is a function of time in these 
epoch, i.e. there is no scaling solution.

We don't plot the ratio $\rho_{\infty}/\rho_{loops}$, where $\rho_{loops}$ 
is the energy density in loops, because 
we arrive the same conclusions as Martins for the open model and the 
Einstein--de Sitter model. In the other two models the strings will never 
dominate the energy density of the universe, first because in models with a 
cosmological constant $L$ increases more rapidly than in the open model, and 
second the $\lambda$--Term approaches 1 for $t \rightarrow \infty$, a value 
which could never be reached by $\rho_{\infty}$ or $\rho_{loops}$.

\begin{figure}[htb] 
  \begin{center}
    \leavevmode
    \epsfxsize=15cm
    \epsffile{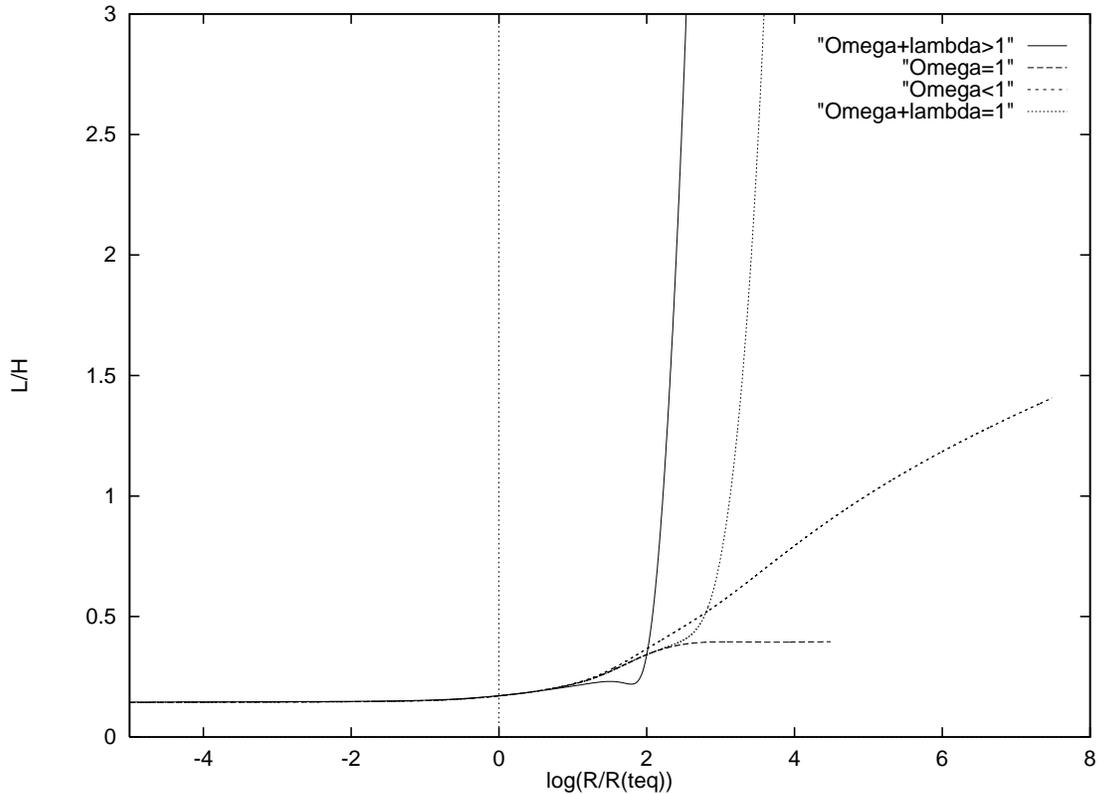}\vspace{0.5cm}
    \caption
    {Ratio $L/H$ as a function of log($R/R_{eq}$).}
    \label{fi2}
  \end{center}
\end{figure}

\begin{figure}[htb] 
  \begin{center}
    \leavevmode
    \epsfxsize=15cm
    \epsffile{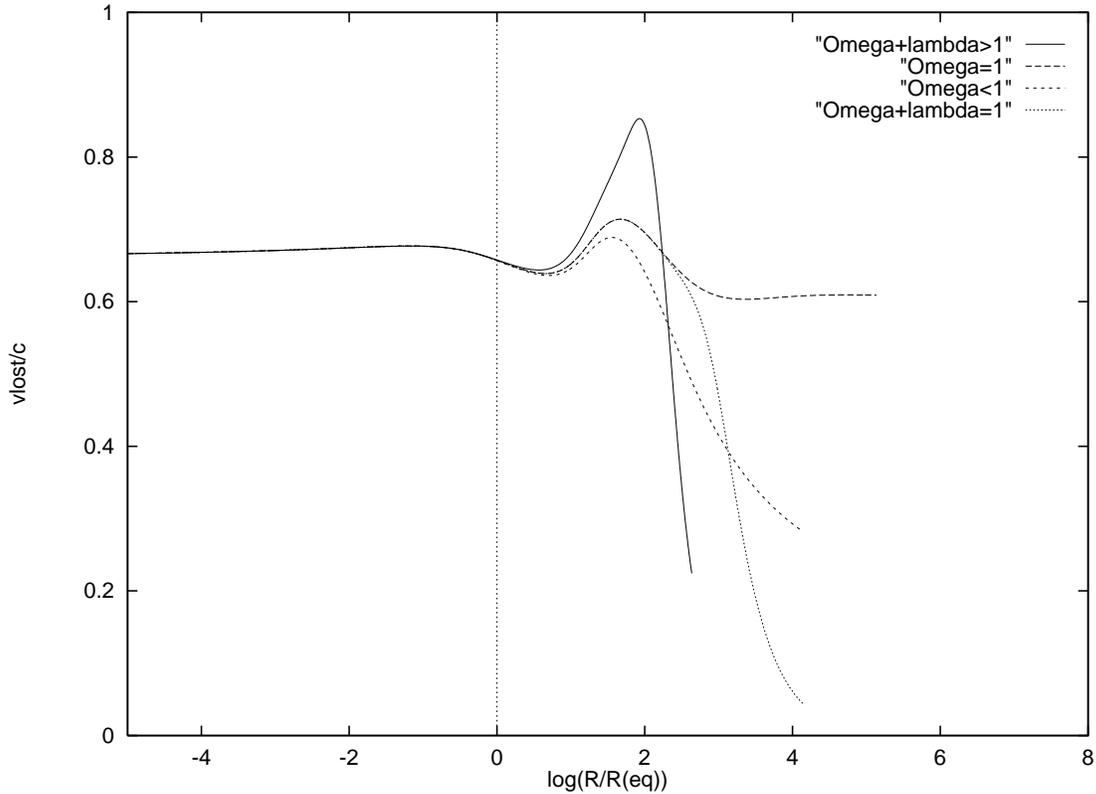}\vspace{0.5cm}
    \caption
    {Long string RMS velocity in units of $c$ as a function of log($R/R_{eq}$).}
    \label{fi3}
  \end{center}
\end{figure}

\begin{figure}[htb] 
  \begin{center}
    \leavevmode
    \epsfxsize=15cm
    \epsffile{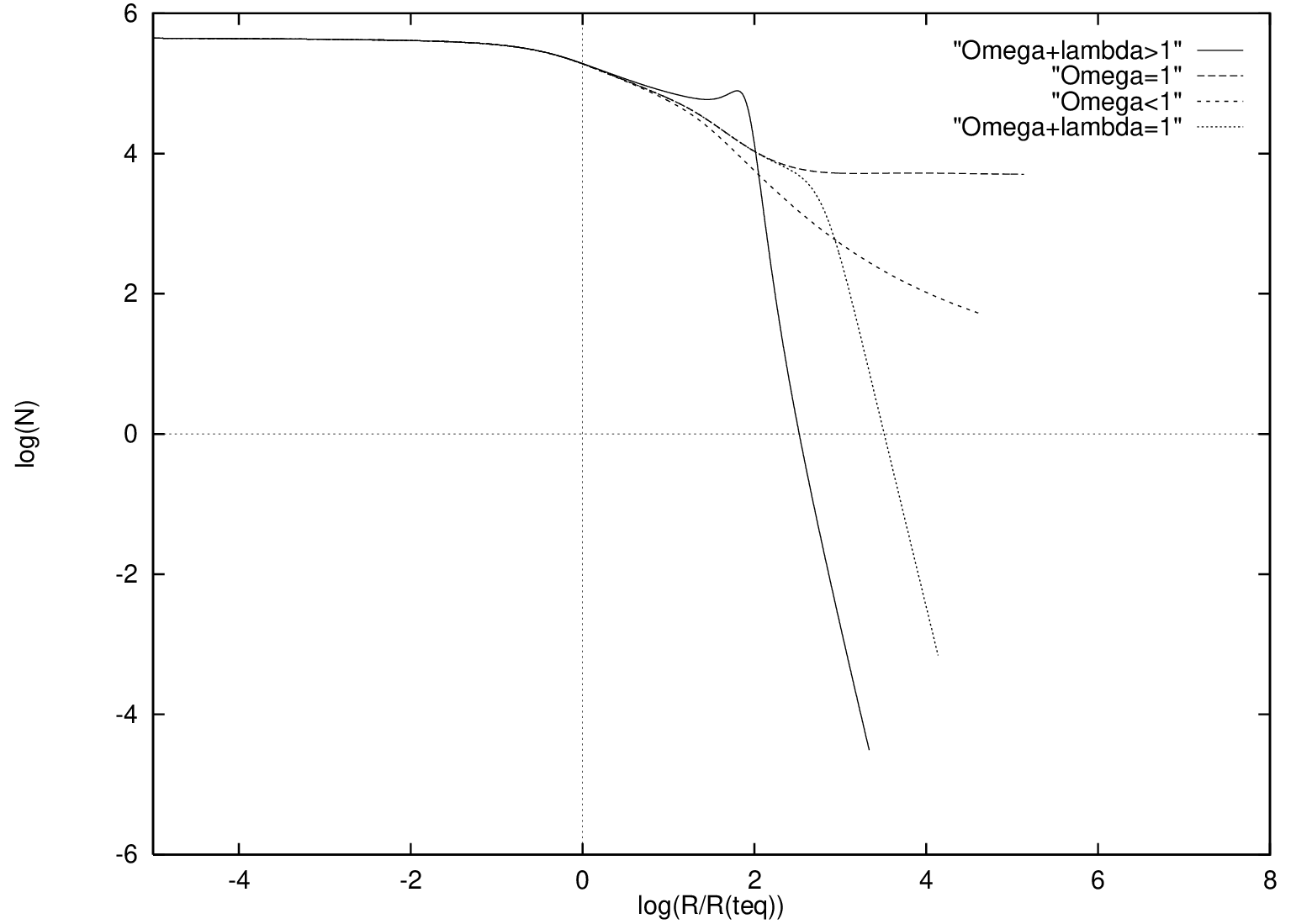}\vspace{0.5cm}
    \caption
    {The logarithm of the number $N$ of loops produced per Hubble volume and 
      Hubble time as a function of log($R/R_{eq}$).}
    \label{fi4}
  \end{center}
\end{figure}



\section{Discussion}
The fact, that there is no scaling solution in the general case has 
important consequences on the structure formation theory with 
cosmic strings. In open universes one expect differences (compared 
to the Einstein--de Sitter model) only on large scales. This could 
have important consequences on the normalisation of the string mass 
per unit length $\mu$ from the COBE data. The same holds for the flat 
model with $\Omega_{0} + \lambda_{0} = 1$. We expect significant 
consequences in the closed model with $\Omega_{0} + \lambda_{0} > 1$. This is
due to the fact, that the loop production rate is higher than in the other 
models (see Figure 4). Thus, loops could play an important role in 
structure formation in this model. One can also see, that the RMS velocity 
is high, suggesting, that the wiggly strings produce wakes rather than 
filaments. If our results can solve the problem of strucutre formation 
with cosmic strings \cite{Pen}, should be investigated in more detail. 

Our results have not only consequences on structure formation theory with 
cosmic strings. The prediction of the gravitational wave background 
and the spectrum of high--energy particles depends also on the network 
evolution. 

Our work based on the ``velicity--dependent'' one--scale model 
by Martins \& Shellard. If this model can describe all transition 
regimes (for example from matter to vacuum regimes) and if the 
ansatz for $k$ is correct will be investigated in more detail in 
future publications. 

Structure formation and the anisotropies in the CMBR due to long cosmic 
strings in these cosmological models are investigated our future work. 

\vspace{1cm}
{\bf Acknowledgements}
The author thanks C.J.A.P. Martins for helpful discussions on the 
velocity--dependent ``one--scale" model. He also thanks W. Priester for his 
careful reading of the original manuscript. This work was supported 
by the Deutsche Forschungsgemeinschaft (DFG).

\end{document}